\documentclass[useAMS,usenatbib]{mn2e}

\usepackage{graphicx}

\title[ANNs for centroiding elongated spots in SH-WFS]{Artificial neural networks for centroiding elongated spots in Shack-Hartmann wavefront sensors}
\author[A. T. Mello et al.]{A. T. Mello$^{1}$, A. Kanaan$^{1}$, D. Guzman$^{2}$ and A. Guesalaga$^{2}$\\
$^{1}$Dept. of Physics, Universidade Federal de Santa Catarina, Campus Universit\'{a}rio Reitor Jo\~{a}o David Ferreira Lima, Florian\'{o}polis, Brazil\\
$^{2}$Dept. of Electrical Engineering, Pontificia Universidad Catolica de Chile, Vicu\~{n}a Mackenna 4860, Santiago, Chile;}
\begin{document}

\date{Accepted 2014 March 3. Received 2014 March 1; in original form 2014 January 9}

\pagerange{\pageref{firstpage}--\pageref{lastpage}} \pubyear{2013}

\maketitle

\label{firstpage}

\begin{abstract}
The use of Adaptive Optics in Extremely Large Telescopes brings new challenges, one of which is the treatment of Shack-Hartmann Wavefront sensors images. When using this type of sensors in conjunction with laser guide stars for sampling the pupil of telescopes with 30+$\,$m in diameter, it is necessary to compute the centroid of elongated spots, whose elongation angle and aspect ratio are changing across the telescope pupil. Existing techniques such as Matched Filter have been considered as the best technique to compute the centroid of elongated spots, however they are not good at coping with the effect of a variation in the Sodium profile. In this work we propose a new technique using artificial neural networks, which take advantage of the neural network's ability to cope with changing conditions, outperforming existing techniques in this context. We have developed comprehensive simulations to explore this technique and compare it with existing algorithms. 

\end{abstract}


\section{Introduction}

The most compact image formed by a telescope is limited by the diffraction pattern known as the Airy disk.  However, owing to atmospheric turbulence, ground based telescopes are generally far from reaching such limit.  The diffraction spot in the visible for a 1$\,$m telescope is 0.117 arcsec FWHM, for a 4$\,$m telescope it is 0.029 arcsec and for a 40$\,$m telescope 0.0029 arcsec.  
At the very best astronomical sites under extraordinary conditions the image of a point source can reach values as small as 0.25 arcsec \citep{Racine1995}, similar to the diffraction figure of a 0.5$\,$m telescope, and much larger than the diffraction limit of larger telescopes.

Adaptive optics (AO), first suggested by \citet{Babcock1953} and first implemented in the 1980s \citep{Merkle1989} partially corrects the wavefront  distorted by atmospheric turbulence.  The rectification of the wavefront is achieved by measuring the wavefront shape and introducing compensating distortions using a deformable mirror.  Measurement and compensation must happen at a time interval shorter than the characteristic time scale for changes in the atmosphere. The typical frequency for such compensations for classical AO systems is around 50 to 250$\,$Hz \citep{Hardy1998}, but modern systems are being planed that can approach 1000$\,$Hz \citep{Davies2012}. The increase in frame rate reflects a need to better sample the turbulence in systems with a higher order of correction.

Current adaptive optics systems deliver angular resolutions down to 22$\,$mas in observations of the Sun, asteroids, atmosphere of planets in the solar system, circumstellar disks, the Galactic centre, and spatially resolving galaxies at z of 1.5 to 3 \citep{Davies2012}.  Some of the key projects for the new generation of extremely large telescopes with 30 to 40$\,$m in diameter are dependent upon the implementation of AO systems on these telescopes. These include direct imaging of exoplanets; resolving stellar populations in nearby galaxies in order to trace their star-formation history and measuring the proper motion of stars in clusters to derive their internal kinematics.

\subsection{The Shack-Hartmann wavefront sensor}
\begin{figure}
  \includegraphics[width=84mm]{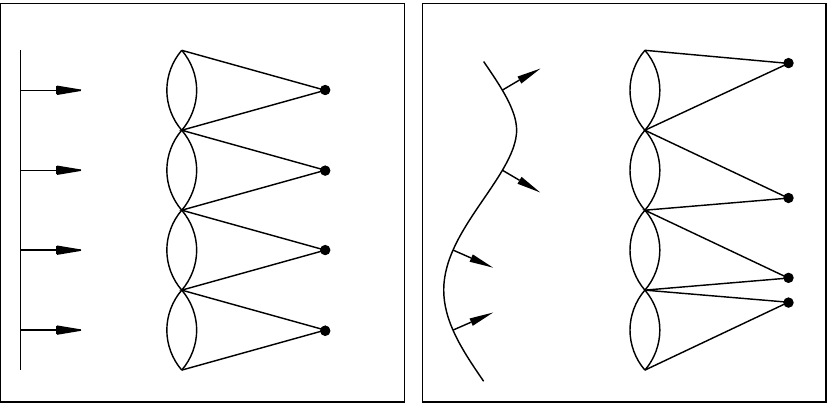}
  \caption{Shack-Hartmann Wavefront Sensor: Left panel shows a plane  wavefront. Right panel shows a distorted wavefront and the corresponding displacement of the spots.}
  \label{figure:SH}
\end{figure}

The Shack-Hartmann wavefront sensor (SH-WFS) consists of an array of lenses used to create multiple of images of a point source, normally a star, onto an image sensor such as a CCD.  Each lens in the array constitutes a subaperture. Fig. \ref{figure:SH} shows a side view of the array. A plane wavefront produces diffraction-limited spots in the projection of each subaperture on the detector, whereas a distorted wavefront displaces and degrades the spots within each subaperture image.

Measuring the spot positions in Fig. \ref{figure:SH} the average wavefront slope at each subaperture can be determined. The AO system will reconstruct the shape of the distorted wavefront from these measurements. Therefore, it is essential to measure the position of the spot centre accurately.

\subsection{Elongated spots}

\begin{figure}
 \includegraphics[width=84mm]{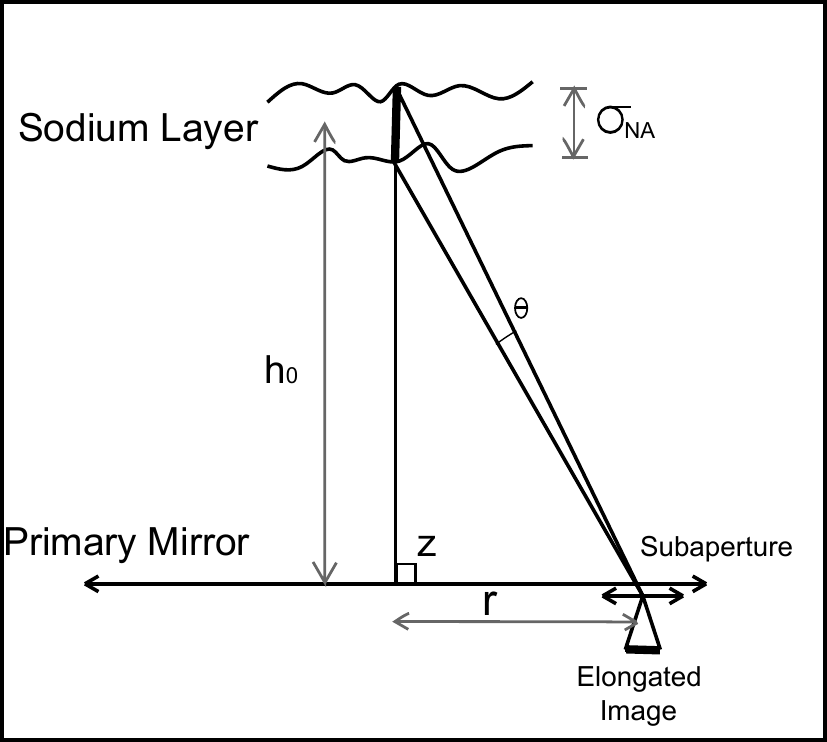}
 \caption{Elongation of Shack-Hartmann spots: an outer subaperture of a SH-WFS is represented.  The artificial star is imaged by the outermost subaperture as an elongated spot.  The image created by the central subapertures is a normal circular spot. $r$ is the distance of the subaperture in the Shack Hartmann from the centre (as projected onto the primary mirror), $h_{0}$ is the sodium layer average altitude, $\sigma_{\it NA}$ is the sodium layer thickness and $z$ is the zenith angle.}
 \label{figure:elongation}
\end{figure}

When there is no suitable star near the observed object, it is necessary to create an artificial star to serve as a reference for the wavefront sensor. There is a layer of sodium atoms at an altitude of approximately 90$\,$km that can be used to produce this artificial star by using a laser focused at this layer. The laser must have a wavelength of 589$\,$nm (D-line of sodium) to excite the sodium atoms that in turn will re-emit the light in this wavelength to provide the return signal. As this layer has a finite thickness of approximately 10$\,$km, the region of excited sodium atoms has a roughly cylindrical shape.  When viewed from the centre of the SH-WFS  this cylinder appears as a spot. When viewed from the periphery of the SH-WFS it becomes an elongated spot.  In 
extremely large telescopes the image of this artificial star created by the outermost subapertures in a Shack-Hartmann is elongated in comparison to the central spot.  Fig.  \ref{figure:elongation} shows a diagram of how this elongation occurs.

In Fig.  \ref{figure:elongation} the primary mirror is schematically represented while highlighting the outermost subaperture of the SH-WFS as projected onto the primary mirror. This subaperture forms the most elongated spot. The elongation depends on the projected distance from centre of telescope pupil and on the altitude and thickness of the sodium layer. The elongation of the spot is approximately \citep{Lardiere2008}:
\begin{equation}
   \theta=\frac {r\cdot \sigma_{\it NA}}{{h_{0}}^{2}}\cos \left( z \right) 
\end{equation}

where $r$ is the distance of the subaperture in the Shack Hartmann from the centre as projected onto the telescope pupil (for maximum elongation it is the primary mirror radius), $h_{0}$ is the sodium layer mean altitude, $\sigma_{\it NA}$ is the sodium layer thickness and $z$ is the zenith angle. For the Thirty Meter Telescope (TMT), \citep{Nelson2006} with $r$ = 15$\,$m, $h_{0}$ = 90$\,$km and $\sigma_{\it NA}$ = 10$\,$km the maximum elongation would be $\theta$=3.82$\,$arcsec at zenith. With a pixel scale of 0.5$\,$arcsec per pixel this gives a 7.64 pixels spot. In this work we are using TMT sized telescope for consistency with the cited references, but the results can also be scaled to other extremely large telescopes.

\begin{figure}
 \includegraphics[width=84mm]{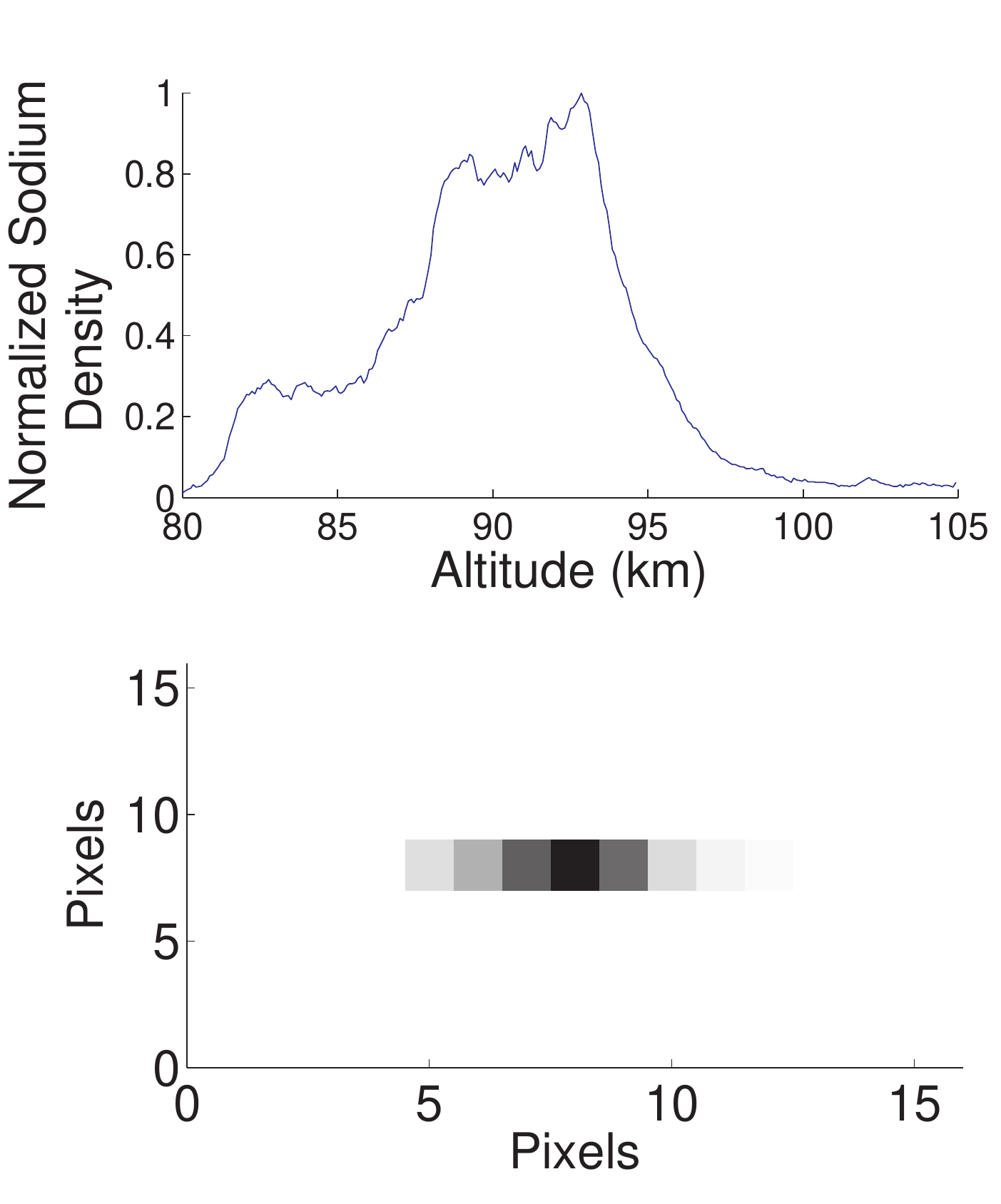}
 \caption{Sodium layer density profile and the corresponding elongated spot. The line-of-sight profile of the sodium density variations are reproduced as intensity variations of the elongated spot.}
 \label{figure:SpotFromProfile}
\end{figure}

The resulting elongated spot is not uniform because of the density variations in the sodium layer with altitude.  In Fig. \ref{figure:SpotFromProfile} an example profile of the sodium layer density is shown, with the resulting elongated spot. These examples come from the measurements done at the Large Zenith Telescope (LZT) using LIDAR \citep{Pfrommer2010}. The density profile also varies in time, so the profile is not stable.

\subsection{Centroiding algorithms}
A centroiding algorithm is a technique to determine the real centre of a spot image. The unpredictable variations in spot shape caused by temporal changes in the sodium layer profile generate elongated spots of varying shapes whose real geometric centre is difficult to determine.

The presence of noise, from photon noise, CCD readout noise, and background light introduce errors in centroid  determination that must be taken into account. A good centroiding technique should be relatively immune from these sources of noise. An analysis of noise and spot elongation influences on centroiding error is provided by \citet{Thomas2008}

Some techniques have been proposed to cope with these problems, the most prominent of which being the matched filter \citep{Gilles2006}, and its improved version, the constrained matched filter \citep{Gilles2008}. The matched filter technique needs a reference image with the shape of the spot being detected, and also a dither signal. 

The dither signal is used as a means of \textquoteleft calibrating' the spot movement for the matched filter. It is made by moving the spot by a known amplitude in the four cardinal directions. Each direction will generate an image that will be used to calibrate the matched filter gain in that direction.
This can be accomplished with a tip-tilt mirror moving the spot away from the center on the Shack-Hartmann subaperture. Both the reference and the dither are required be updated frequently to keep up with the changes in the sodium profile which affects the shape of the elongated spot.

Another technique used determining the centroids of elongated spots is correlation tracking \citep{Michau2006}. This technique uses a reference model of the elongated spot and correlates this with the spot image to determine its centre. This technique requires that the current sodium density profile be known for the reference.

In this paper we present results on the use of artificial neural networks (ANNs) to identify the correct centre of the spot in the presence of noise and elongation.
This technique does not need the use of reference or dither.

\subsubsection{Constrained matched filter implementation}

For comparison with our ANN technique, we also implemented the centre of gravity (CoG) and the constrained matched filter techniques as a reference. Information on the implementation of the centre of gravity techniques can be found at \citet{Thomas2006}.
The Centre of Gravity technique, although not appropriate for elongated spots, was implemented to establish a reference point to judge how much is gained with the other techniques. We chose not to compare our system with the correlation tracking technique because it uses a reference with arbitrary resolution that needs to be optimized, and we decided to concentrate our work in only one technique, the constrained matched filter. \citet{Lardiere2010} shows that the constrained matched filter gives very close, and sometimes better results than correlation tracking.

The implementation of the constrained matched filter in this work follows exactly that described by \citet{Gilles2008}, but we had to choose the value for the dither displacements as the authors did not specify it. \citet{Lardiere2008} tested several dither values on a variety of conditions, 0.02 pixels was the best and we verified that with our data, so it was adopted.
In our simulation the reference image was constructed using the average of the last 5 images to create a better SN version of the elongated spot. The dithered images were also averaged for better SN. The dither and reference signals are always updated in this work, this is impractical in reality but gives the best possible results. The Matched Filter technique uses a linear filter that is noise-weighted, and we used the image being processed in each iteration to extract the noise vector.

\subsection{Artificial neural networks}

An artificial neural network (ANN) is a computational system, inspired by the working of the brain, that can be used in complex and non-linear calculations and control systems. The ANN is comprised of a number of nodes, called neurons, connected to inputs and outputs by a weight function. The neuron itself processes the received inputs, normally mathematically summing all the weighted inputs and applying to this value either a linear or a non-linear function, a sigmoid, for example.

Each neuron is connected directly to one output by another weighting function. The ANN can have one or more hidden layers of neurons. A hidden layer is one that is inside the ANN, receiving inputs and outputs only from other neurons, and not from outside the ANN. The outputs of the first layer serve as inputs for the next layer. The number of layers can be chosen by the ANN designer with the objective of obtaining the best results. The weights connecting the neurons represent the ANN “knowledge”, the weight value reflects the importance of the corresponding input to the neuron. To assign the values to all the weights it is necessary to “train” the ANN. Training is done by showing the ANN a set of inputs with its corresponding outputs. To do that a dataset with the known correct outputs is needed. An algorithm is then applied to obtain the desired weights. Although each individual neuron implements its function slowly and imperfectly, the whole structure is capable of learning complex functions and solutions quite efficiently \citep{Reilly1990}.

\begin{figure}
 \includegraphics[width=84mm]{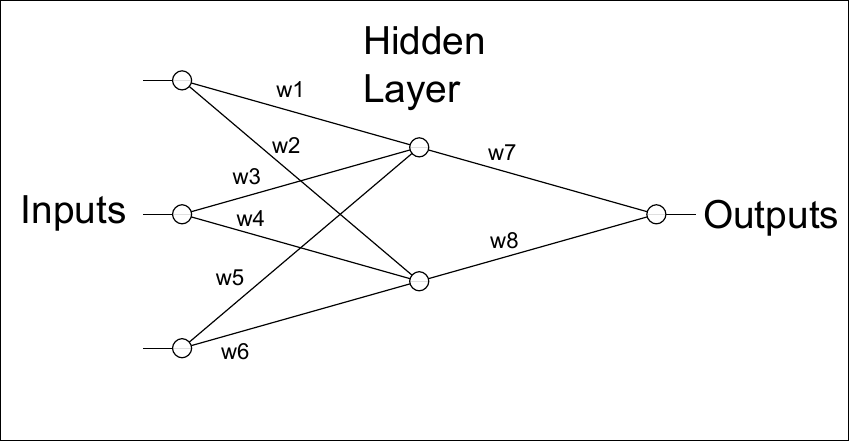}
 \caption{Example of an ANN architecture. This example has three inputs, two neurons in an intermediate layer, called the hidden layer, and one output. The neurons are connected by weights. Each neuron applies a function over the sum of its weighted inputs. The weight values are determined when training an ANN.}
 \label{figure:neuralnetwork}
\end{figure}

Learning algorithms search through the solution space to find a function that has the best possible result. The backpropagation training algorithm, used in this work, attempts to minimize the least mean square difference over the entire training set. The training set is made up of a large number of cases for which the outcome is already known.
Fig. \ref{figure:neuralnetwork} is a schematic diagram of an ANN. This example has three inputs, two neurons in an intermediate layer, called the hidden layer, and one output. The neurons are connected by weights, and these weights are the values that are determined when training an ANN.

It is important to have a large training set so that it has enough variations of scenarios. The ANN can than be trained to be able to cope with all possible scenarios. Once trained the ANN must be validated with data the ANN has not seen during training \citep{Bottaci1997}. This assures the ANN is working and is able to generalize correctly.

It is not possible to predict what is the best ANN topology or sample size, and this needs to be determined by experimentation. Lessons learned from \citet{Osborn2012} using ANN for adaptive optics systems guided our work here. One of them being the use of simulated data to train the ANN to be validated with real data.

\section{Simulations for artificial neural networks}

\begin{figure}
 \includegraphics[width=84mm]{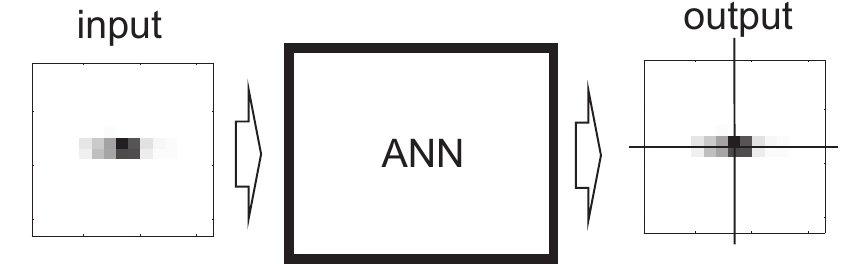}
 \caption{The inputs of the ANN are the subaperture image pixels. The outputs are the x and y position of the real centre. To train the ANN the outputs are provided with the inputs;  when being used, the ANN should generate the correct outputs automatically.}
 \label{figure:ANNinout}
\end{figure}

In the case being described here the inputs for the ANN are the pixel counts from each subaperture of the Shack-Hartmann sensor. The ANN's job consists of determining the centre of the spots formed by each subaperture.   The outputs of the ANN
will be the horizontal and vertical centre of the spot. This is illustrated in Fig. \ref{figure:ANNinout}.

To train the ANNs we used simulated images of the elongated spots. The ANN is presented with pairs of input images and the corresponding output slopes. If presented with a large set of input-output pairs it will learn to recognize the slopes that correspond to each input image. From our experience, for this training to cope with any spot position and profile shape, each training spot image should be computed with a random position and random sodium layer profile. The simulated spot image will use a random profile that is modelled after a real profile.

\begin{figure}
 \includegraphics[width=84mm]{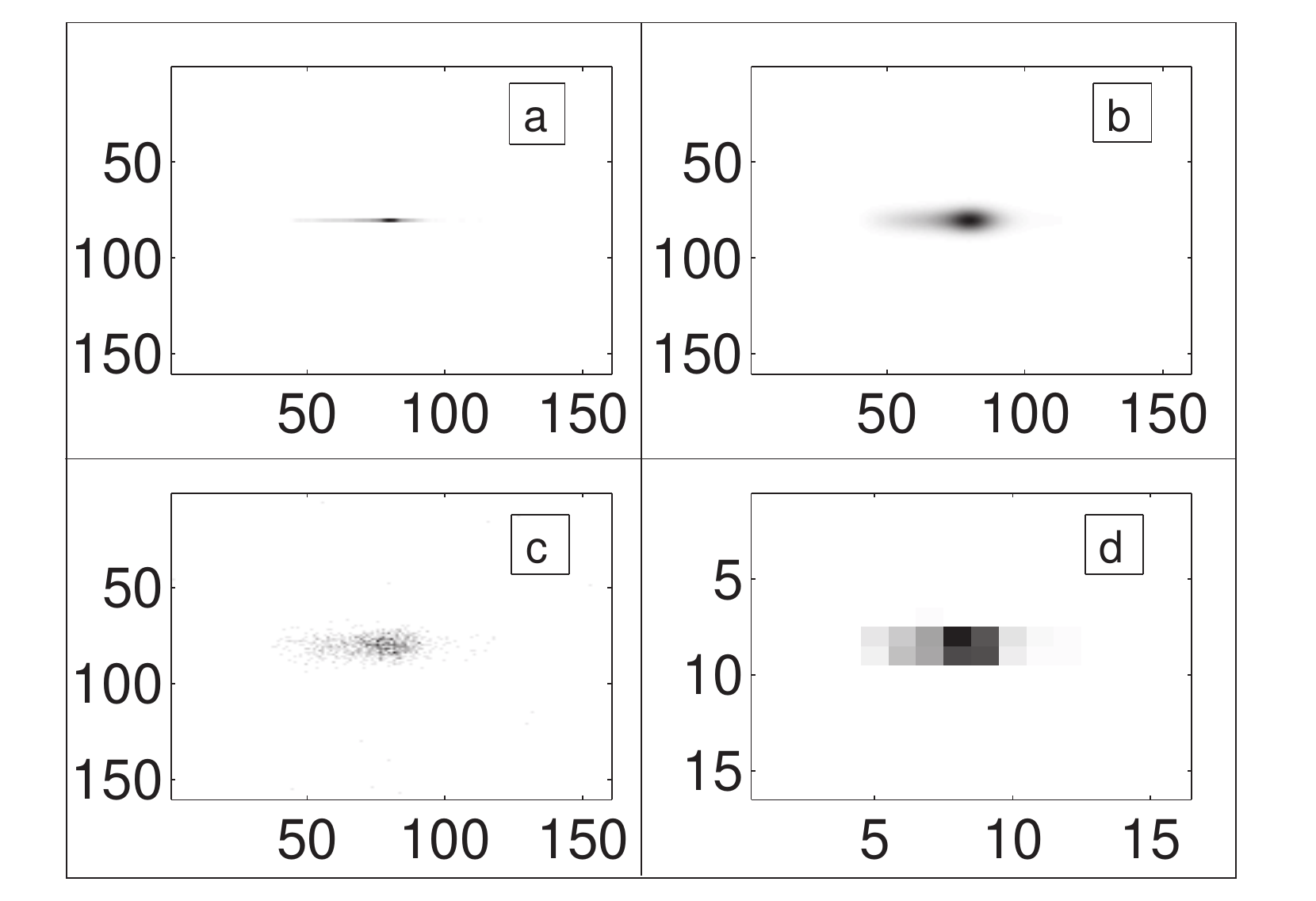}
 \caption{Image simulation for the outermost Shack-Hartmann subaperture, in this case the spot elongation is aligned with a CCD line.  Dimensions are in pixels.  Here we show the simulation steps to form an elongated spot. a: this line represents the sodium profile projection onto the CCD by the subaperture corresponding to this region. b: previous line convolved with a Gaussian to simulate seeing effects. c: photon noise added to the previous spot image. d: re-sampling to the detector pixel scale.}
 \label{figure:subapsim}
\end{figure}

The simulation of the elongated spot, shown in Fig. \ref{figure:subapsim}, is implemented as follows: a line is constructed with the required elongation and sodium layer density profile, as if the artificial star would be imaged by a perfect optical system (panel a). The profile is reproduced in this line by the intensity in each pixel. This line is then convolved with a Gaussian with the same size that an Airy disk would have considering the size of the lens and the pixel scale (panel b). In the case under study the subaperture diameter is 0.5$\,$m, with a pixel scale of 0.5$\,$arcsec per pixel; this results in a spot of 1.19 pixels FWHM. If the Gaussian is not centred the resulting convolved image will also not be centred, and this is used to simulate the required slope.

The resulting image is a long exposure and noise-free elongated spot. Next we add noise. Photon noise is added using the generated image as a template and generating photons with a Monte Carlo simulation (panel c). The Monte Carlo simulation is done as follows: a random pixel in the image is chosen, the template determines the probability of a photon falling on each pixel, so a new random number is chosen to decide if this photon falls or not. It is repeated until the chosen number of photons falls onto the detector. The resulting image is at a higher spatial resolution than the system being modelled, so we lower the resolution of the image to the 16x16 pixels used for the subaperture in this work (panel d).

For the spot position in the training set, the spot is positioned randomly following a normal distribution. As a result, the training set will have the spot positioned anywhere around the sensor, but with a much higher probability of being near the centre. This is closely related to what the ANN will see when the image movement is created by turbulence with Kolmogorov statistics.

\subsection{Profile modelling}

We avoided using real profiles of the sodium layer density in training the ANN. Simulated profiles can be much more varied so that the ANN can cope with cases not seen in a limited set of real measurements. Previous work with ANNs \citep{Osborn2012} have shown that the use of simulated data with the same statistics as the real data in the training set
gives results close to the ones trained with real data, with the advantage of having more diversity in the training set.

\begin{figure}
 \includegraphics[width=84mm]{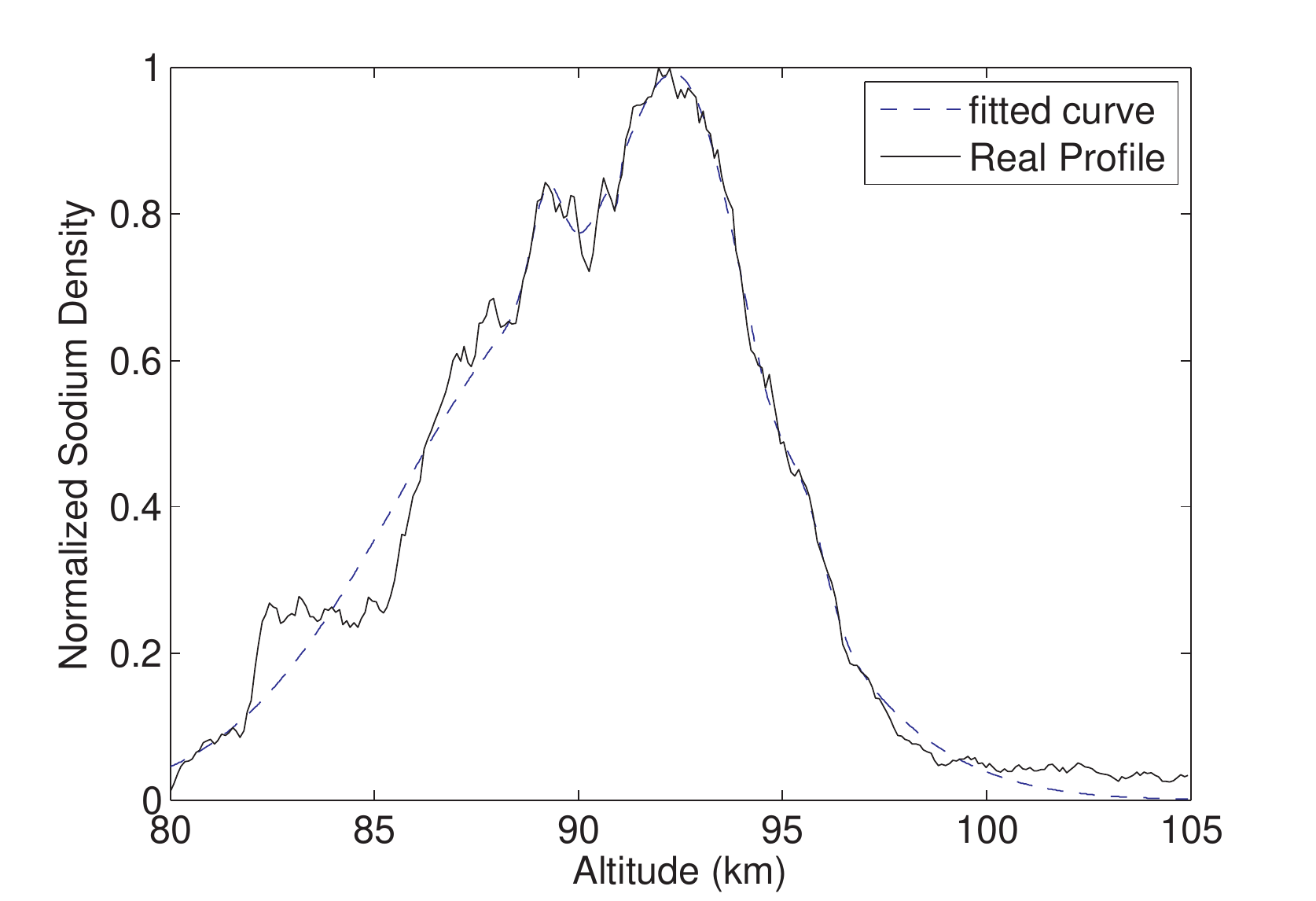}
 \caption{Fitting five Gaussians to a real sodium profile.  The fits are repeated for a large number of profiles.  The statistics of the Gaussian parameters are stored.  New synthetic profiles are created at random using five Gaussians which obey the recovered statistics. }
 \label{figure:ProfileFit}
\end{figure}

For construction of the sodium layer model we used measurements obtained by the LIDAR facility of the LZT \citep{Pfrommer2010}. Each real measurement of the sodium layer density profile was fitted with five Gaussians. An example of a fitted profile can be seen in Fig. \ref{figure:ProfileFit}.

Each Gaussian can be constructed with three parameters, according to
\begin{equation}
   f(x) = ae^{-(x-b)^2/c^2 }
\end{equation}

\begin{figure}
 \includegraphics[width=84mm]{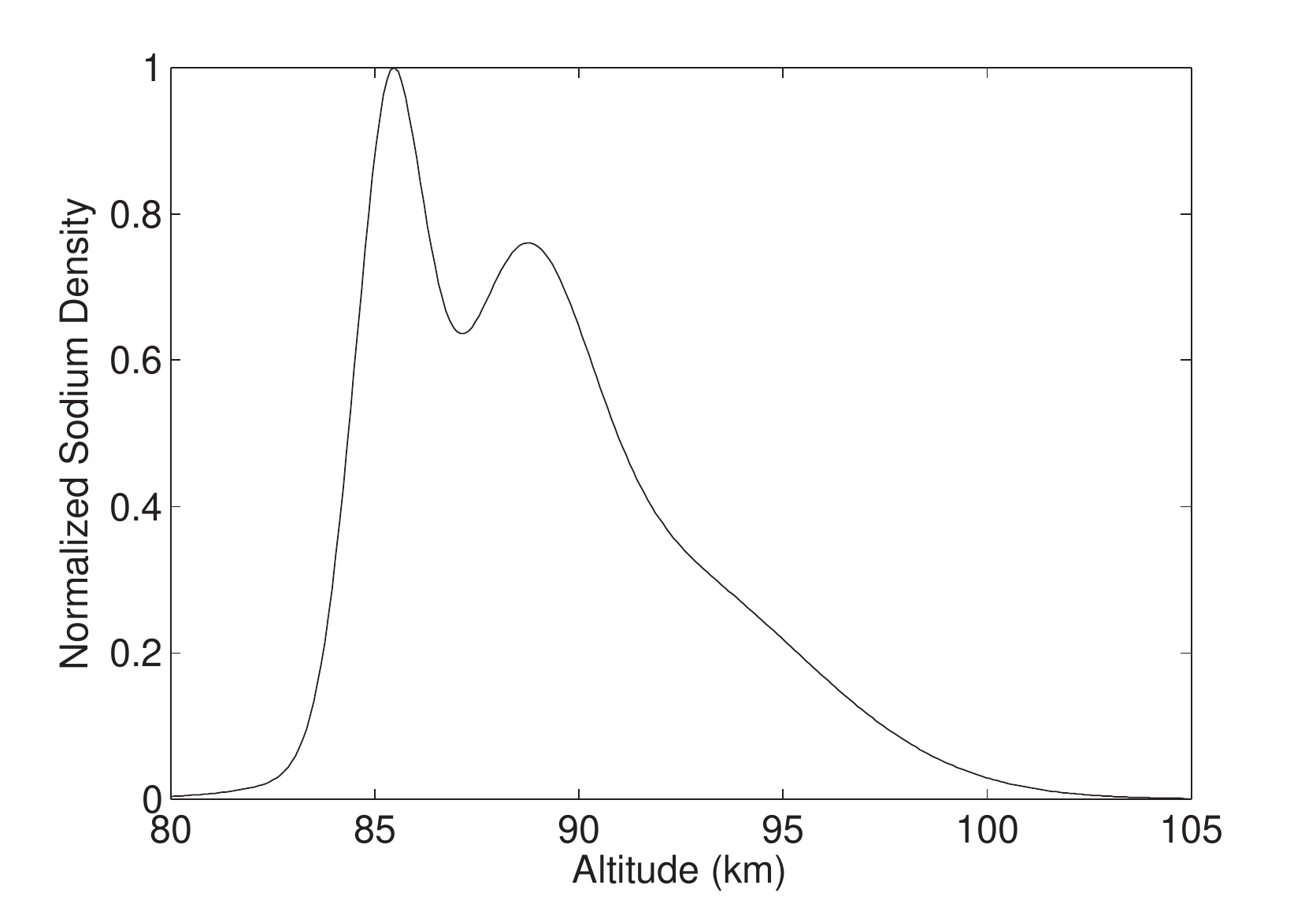}
 \caption{Example of a simulated Sodium Profile using five Gaussians as explained in Fig.  \ref{figure:ProfileFit}.}
 \label{figure:ProfileSim}
\end{figure}

The parameters, $a$, $b$ and $c$, for each of the five fitted Gaussians follow a statistical distribution. We characterize this distribution of the parameters as a normal distribution and determine the mean and the variance for each of the five Gaussians.  These variances and means are used to create a Monte Carlo simulation to generate synthetic profiles which are the sum of the five Gaussians. To train the ANN we typically create one million synthetic profiles.  An example of a single simulated profile constructed using this method is shown in Fig. \ref{figure:ProfileSim}.

Simpler models of sodium layer profiles were also tried. With less than five Gaussians the trained ANN results were unacceptable, the average resulting error was bigger than the ANN trained with five Gaussians. For six or more Gaussians there was no appreciable gain. More details on how the error was calculated are provided in section \ref{section:AvErr}

\subsection{Artificial neural networks}

Because in a normal Shack-Hartmann image elongation of the spots may not be aligned with the CCD lines, the use of a special CCD, a polar coordinate detector, is being developed to cope with elongated spots \citep{Adkins2006}. A more complete description of this detector can be found in \citet{Adkins2012}. This polar coordinate detector would have the pixels always aligned with the elongation of the spots. The constrained matched filter technique was developed for this detector type so we developed some ANNs designed to work in this case. But we are also interested in the use of ANNs in normal types of CCDs, where the elongation is not aligned with the detector pixels, and some of our results pertain to this. To differentiate it from the polar coordinate detector we are going to call it the cartesian CCD. The limiting factor of the cartesian CCD is that the ANNs do not manage different elongation directions well, enforcing the need to train a different ANN for each direction of elongation. The advantage is that ANNs so trained still work with cartesian CCDs, allowing the use of conventional CCDs.

\subsection{Validation}

To validate a previously trained ANN we expose the ANN to a dataset which was not used during training.  In our case we trained the ANN with simulated data following the statistics of LIDAR observations of the sodium layer.  To validate the ANN  we used real sodium profiles from the LZT LIDAR experiment \citep{Pfrommer2010} to provide the asymmetries in the spot.  To create spot movement on the simulated Shack-Hartmann we created Kolmogorov turbulence on a virtual phase-screen.

To simulate a time evolving sodium profile the LZT LIDAR data needed interpolation.  LZT LIDAR data have a time resolution of one second, while we need to simulate a system running at 700$\,$Hz, the frequency a real adaptive optics system of this type would operate at.  Therefore linear interpolation was used to obtain a continuously evolving profile.

When training the ANN, wavefront slopes were created randomly and time independently. But for the validation data, time dependent slopes were obtained by generating a phase-screen with Kolmogorov statistics and moving it with a speed of 10$\,\rmn{m s^{-1}}$.

\subsection{Artificial neural network architecture}

All ANNs used in this work have 256 inputs, representing the 16x16 pixels of the subaperture image, and 2 outputs representing the horizontal and vertical spot displacements. For the hidden layer, 16 neurons were used, and the training method used was the Levenberg-Marquardt backpropagation. \citep{Marquardt1963}

Our ANNs were trained using Matlab. Matlab works well in a simulated environment but for a real system it would be difficult to port it to a local controller, and as a scripting language Matlab’s performance would be much worse than other compiled languages like C or C++.

Once an ANN is trained and  validated it can be implemented on a local controller using a system like DARC \citep{Basden2012} which is an open source high performance real-time control system for astronomical adaptive optics systems.

\section{Results}

In this section we compare the ANN results obtained in the validation tests with other centroiding techniques: centre of gravity and the constrained matched filter.

\subsection{Average error} \label{section:AvErr}

To evaluate the performance of our algorithms we use centroiding error as a figure of merit. To test a centroiding algorithm we simulated a subaperture image with given elongation and position. We then compare the simulated position for the spot with the position measured by the centroiding algorithm. 
The centroiding error is the absolute difference between the centre for the simulated spot and the value measured by the algorithm.

We experimented with the biggest elongation in the spot as a worst case scenario. We carefully implemented the existing centroiding techniques for direct comparison with our proposed algorithm.

A complete SH-WFS with a number of subapertures is required to produce results in terms of wavefront error. As we are introducing the technique, we do not consider wavefront error results necessary at this stage.  Instead, we use centroiding error as a figure of merit, since the smaller the centroiding error, the lower the WFE will be. To test how the ANNs cope with varying levels of noise we designed a set of tests in a range of noise levels; in this case photon noise, which vary from 100 to 30000 photons.

The noisy images were generated using only photon noise for simplicity, so this corresponds to signal to noise (SN) 
ranging from 10 to 173. This brackets the expected photon throughput for the E-ELT using laser guide stars, varying from 300 - 1100 photons per pixel per frame \citep{Lardiere2010}. Although we are using TMT as a reference in this paper, we have no information on expected photon levels for it so in this case we are using the information available for the E-ELT.

Ten thousand iterations were computed for each noise level; the results presented in Figs. \ref{figure:avgideal} to \ref{figure:avgdiag} are the average centroiding error for each technique for all the iterations. The average centroiding error was computed as the average throughout all iterations of the absolute error; the absolute error being the absolute value of the difference between the measured centre and the real centre. These values are then compared with other centroiding techniques.

Next we show four different cases. In these we used two different ANNs. The first was trained in the presence of noise,  chosen to be 1000 photons coinciding with the higher expected value for the E-ELT; this is referred to ANN 1K in the plots.  The second is trained without the presence of noise; this is referred to as ANN Noiseless in the plots.

\subsubsection{Ideal case}

\begin{figure}
 \includegraphics[width=84mm]{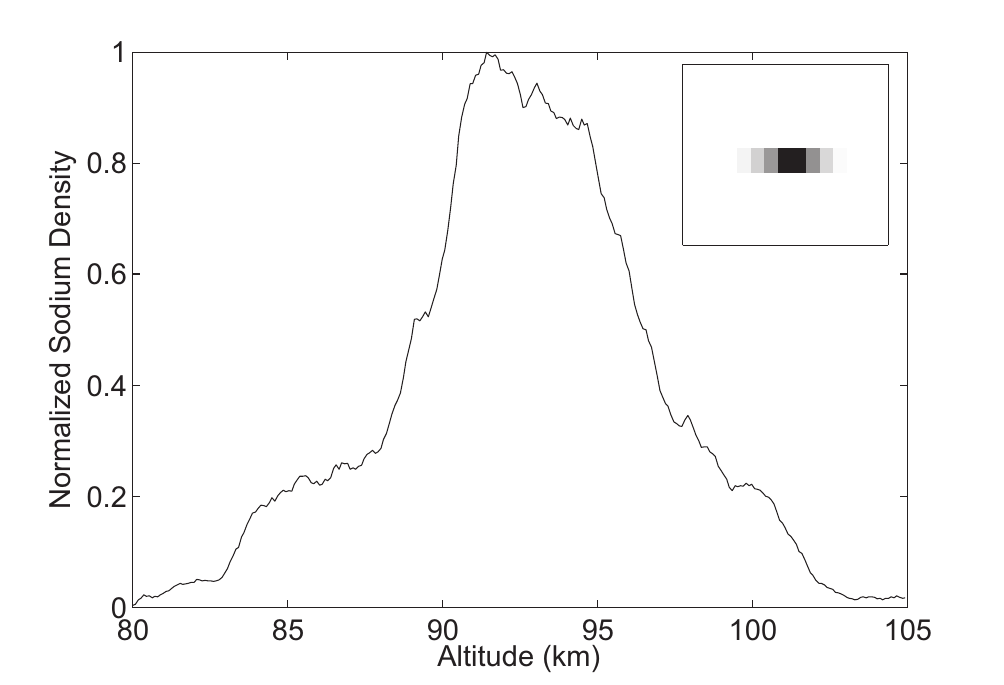}
 \caption{Profile with centred centre of gravity used to test the ideal case. This profile introduces no bias in the centroiding. The inset shows the resulting elongated spot.}
 \label{figure:ProfileCentre}
\end{figure}

To determine the best possible scenario we ran a simulation based on LIDAR measurements where the centre of gravity coincided with the geometric centre of the subaperture. These cases give the best result for CoG.  The simulation had no turbulence included and the detector used was a polar coordinate CCD, the spot elongation of this detector and the employed sodium layer density profile are shown in Fig. \ref{figure:ProfileCentre}. In this simulation we are looking at the effects of photon noise only.

\begin{figure}
 \includegraphics[width=84mm]{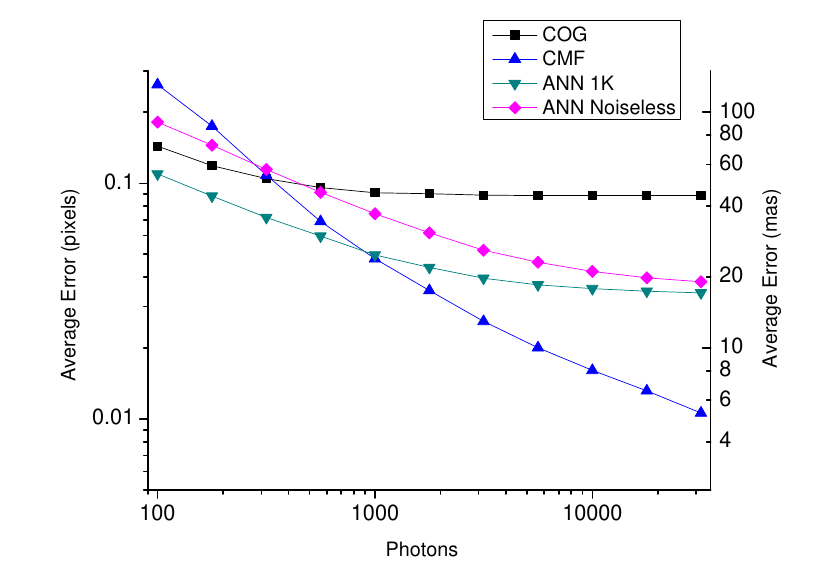}
 \caption{Average pixel error as a function of total detected photons for the ideal case: centred CoG sodium profile in a polar coordinate detector with no turbulence.  COG=centre of gravity method, CMF = constrained matched filter method, ANN 1k = ANN method trained with noise and ANN Noiseless = ANN method trained without noise.}
 \label{figure:avgideal}
\end{figure}

Results for the average centroiding error are shown in Fig. \ref{figure:avgideal} for the CoG, constrained matched filter and ANN Methods.

As it can be seen in the results, the ANN trained with noise (ANN 1K) is better than the ANN trained with no noise (ANN Noiseless) for situations in the presence of high noise. 
As our ANN was trained at a photon level of 1000 photons it operates well in situations with this noise level or higher, but as the noise levels are reduced it stops improving and a plateau is reached. The ANN trained without noise gives worse results at high noise but keeps improving until very low noise situations are reached.

\subsubsection{Uncentred sodium layer CoG case}

\begin{figure}
 \includegraphics[width=84mm]{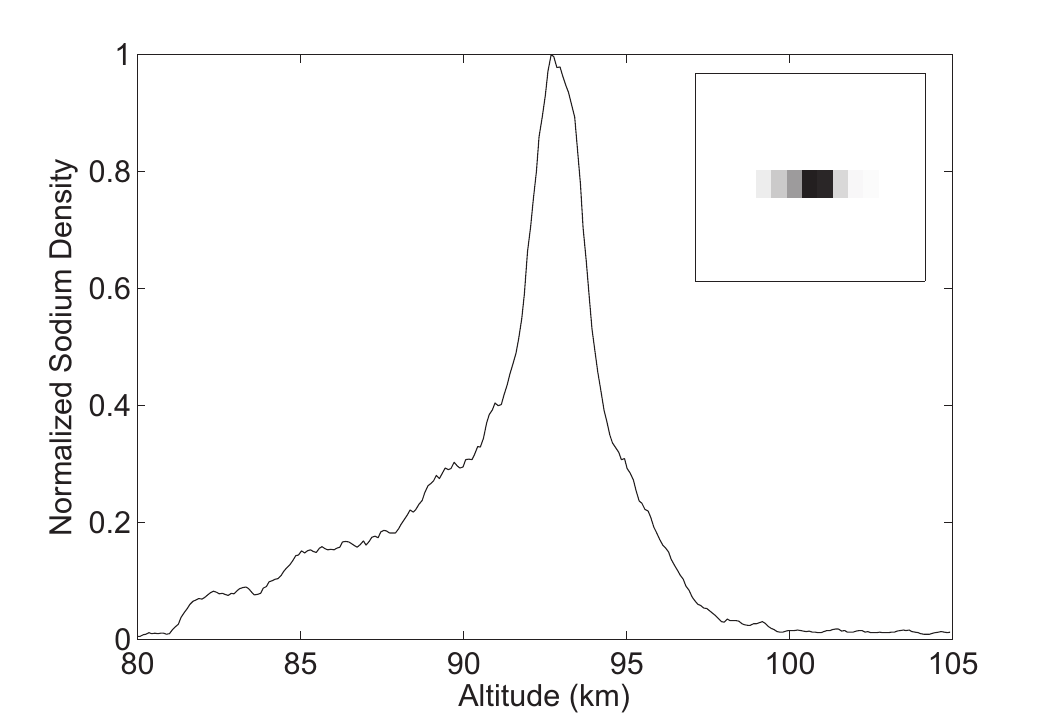}
 \caption{Profile with uncentred centre of gravity used to test the uncentred case and also the turbulent case. This profile introduces bias in the centroiding. In the inset the resulting elongated spot is shown.}
 \label{figure:uncentredprofile}
\end{figure}

In this section we present the results using an uncentred centre of gravity profile, shown in Fig. \ref{figure:uncentredprofile}, polar coordinate detector and no turbulence. This profile has a stronger density on its lower part, from 80$\,$km to 90$\,$km, and this creates a bias for the CoG methods.

\begin{figure}
 \includegraphics[width=84mm]{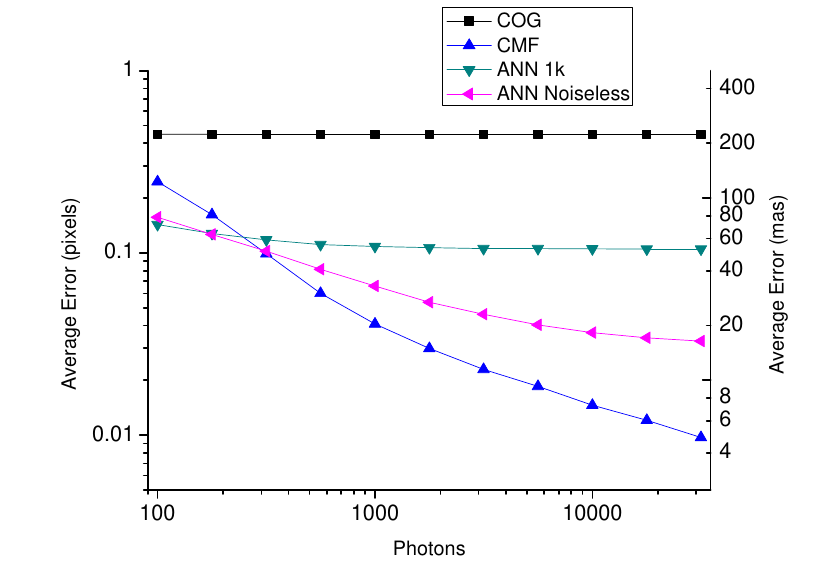}
 \caption{Average pixel error as a function of total detected photons in the uncentred case: uncentred CoG sodium profile in a polar coordinate detector with no turbulence. COG=centre of gravity method, CMF = constrained matched filter method, ANN 1k = ANN method trained with noise and ANN Noiseless = ANN method trained without noise.}
 \label{figure:avgdecent}
\end{figure}

Fig. \ref{figure:avgdecent} shows the average pixel error result for the CoG, constrained matched filter and ANN methods.

As the profile is not centred, the CoG method has a big systematic error, so the error stays around 0.45 pixels.  There is a clear advantage in using the constrained matched filter. Both the matched filter and the ANN trained without noise gets continuously better as the noise is reduced. The ANN trained with noise (ANN 1k) hits a plateau at low noise and does not improve beyond that point.

The constrained matched filter is better in these results, but we should point out that this is an idealized case, there is no turbulence and also the matched filter reference and dither are  updated every frame, which is not feasible in practice. We should expect a better performance from the matched filter in this case because the spot is stationary and the matched filter uses a reference. The spot will be always at the same place as the reference in this unmoving spot case.

Even the method using ANN trained without noise appears better than the one trained with noise, but things change when turbulence is added as the next case shows.

\subsubsection{Turbulent case}

\begin{figure}
  \includegraphics[width=84mm]{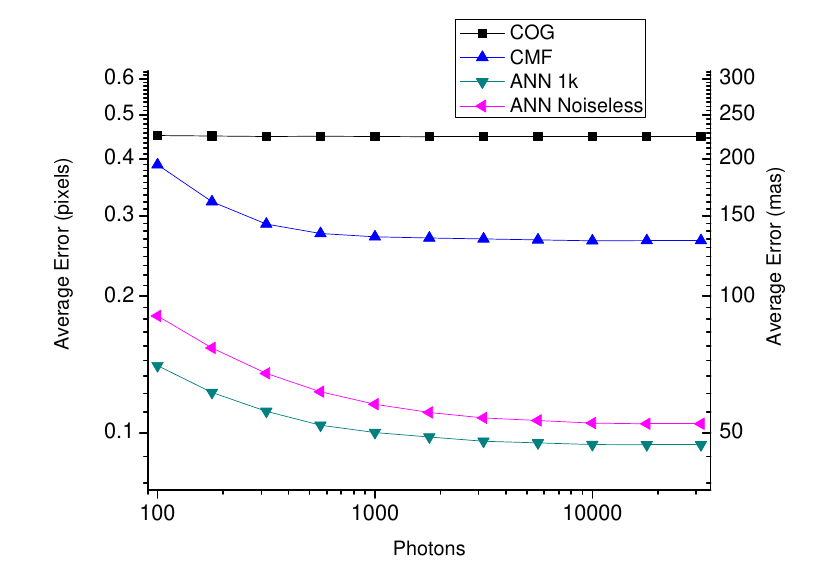}
  \caption{Average pixel error as a function of total detected photons in the turbulent case: uncentred CoG sodium profile in a Polar coordinate detector with the presence of turbulence. COG=centre of gravity method, CMF = constrained matched filter method, ANN 1k = ANN method trained with noise and ANN Noiseless = ANN method trained without noise.}
  \label{figure:avgturb}
\end{figure}

Next, the same tests are made in the presence of turbulence, polar coordinate detector and the same uncentred centre of gravity profile shown in the last case (Fig. \ref{figure:uncentredprofile}). The average error is shown in Fig. \ref{figure:avgturb}.

\begin{figure}
 \includegraphics[width=84mm]{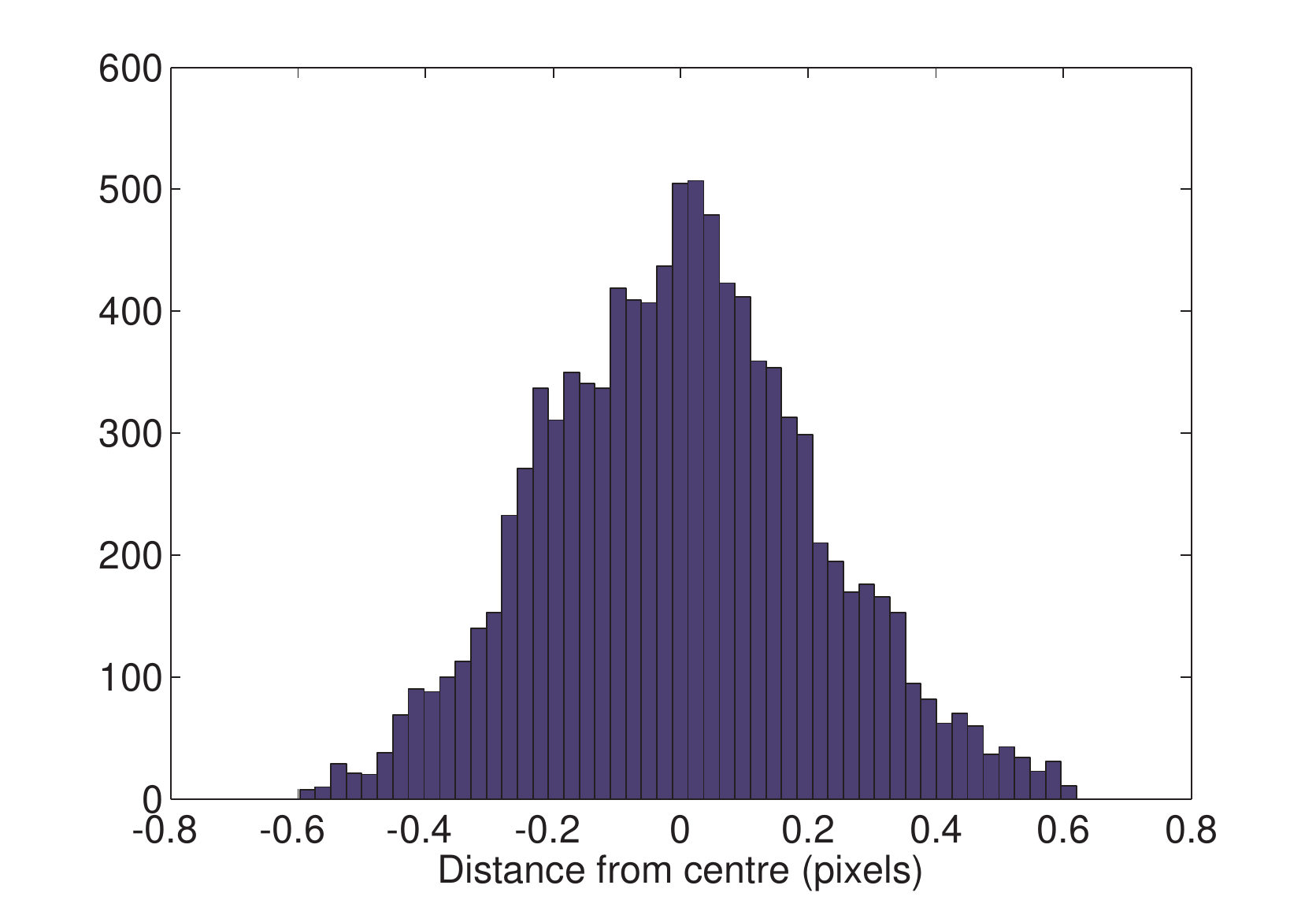}
 \caption{Spot position histogram.  Turbulence moves the spot about the average position.}
 \label{figure:spothist}
\end{figure}

In the presence of turbulence, the ANN method is the best. The turbulence was modelled by phase-screens following Kolmogorov statistics, with a Fried parameter $r_{0} = 0.15\,$m and a static sodium layer profile. This turbulence resulted in a spot movement limited to one pixel, as shown in the histogram in Fig. \ref{figure:spothist}.

The constrained matched filter method does not perform well in the presence of turbulence. It should be expected in reference to the previous cases, for it to be better since it uses a reference that is in the exact same place as the current spot, making it the optimum case for spot centroiding. But in the case with turbulence the spot is moving and the reference will be shifted, making the centroiding more unstable and error prone for this technique. The ANN method uses no references and the spot movement has no bearing on the centroiding.

\subsubsection{Cartesian CCD case with diagonal elongation}

\begin{figure}
  \includegraphics[width=84mm]{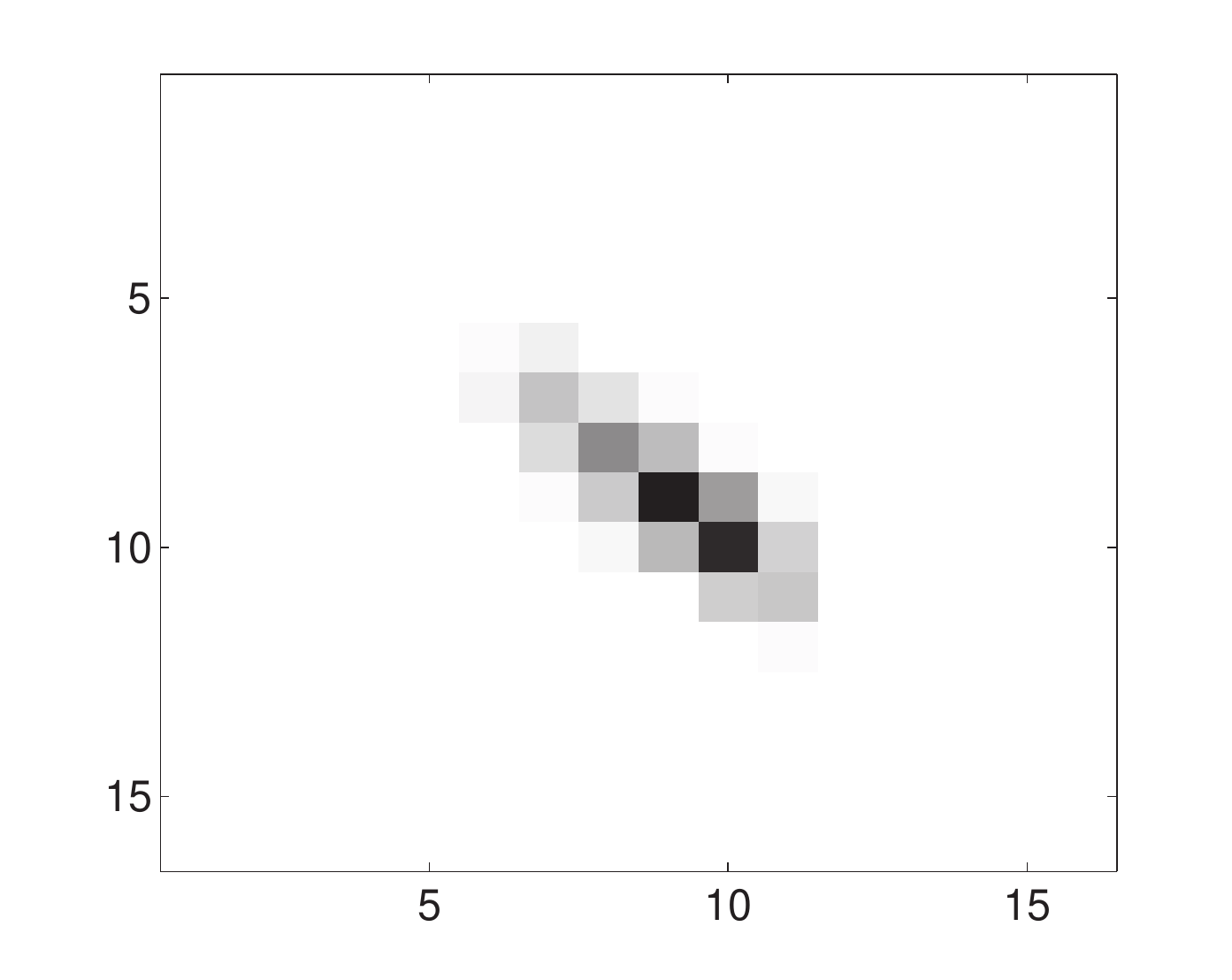}
  \caption{Diagonally elongated spot used to test the techniques in the cartesian CCD case. This is a spot with a diagonal elongation furthest from the centre, providing the most elongation.}
  \label{figure:diagSubap}
\end{figure}

\begin{figure}
  \includegraphics[width=84mm]{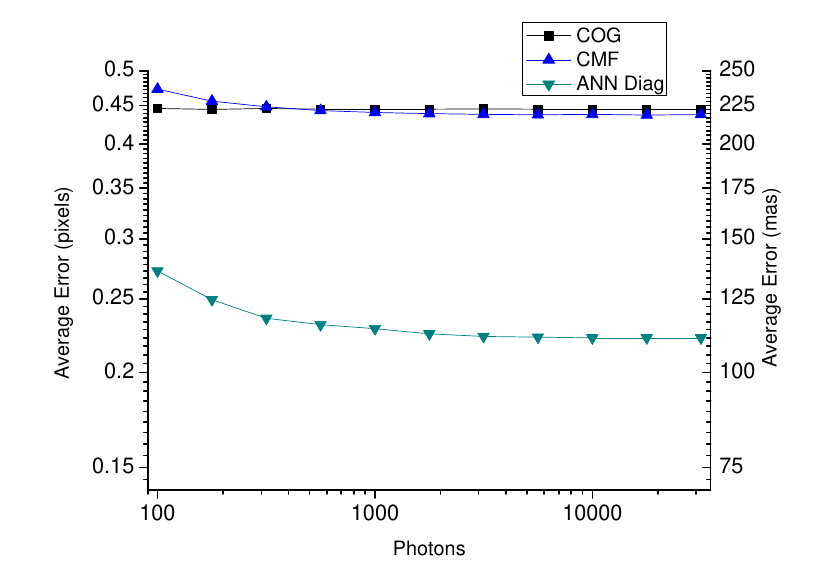}
  \caption{Average pixel error as a function of total detected photons in the diagonal case: uncentred CoG sodium profile in a cartesian CCD with diagonally elongated spots and no turbulence. COG=centre of gravity method, CMF = constrained matched filter method, ANN Diag = ANN method trained with diagonally elongated spots. The ANNs are clearly better than any other method. }
  \label{figure:avgdiag}
\end{figure}

In this section we present results for a conventional cartesian CCD, where the spot elongations are not aligned with CCD lines or columns. Fig.  \ref{figure:diagSubap} shows a subaperture with maximum diagonal elongation. In this simulation the tests are made in the presence of turbulence, with the same uncentred centre of gravity profile shown in the last case, but with a cartesian CCD. We now discuss the results for this situation in the presence of turbulence.  Fig. \ref{figure:avgdiag} shows the average error. The ANN was specially trained for this spot elongation direction, it was trained in the presence of noise at a photon level of 1000 photons and is called in the plot ANN Diag.  For future implementation in a real system, we must train the ANN for each subaperture of the Shack-Hartmann; a computationally expensive task which however, needs to be done only once. As it can be seen, in this case the ANN method gives even better results than other cases.

The above results are summarized in Table \ref{table:ressummary} for 300 and 1000 photons, which correspond to SN of 17.32 and 31.6, the nearest given the expected throughput for the E-ELT. The ANN used in the table is the one that yielded best results, trained with the presence of noise. Table \ref{table:ressummarypercent} shows the same results in percentages.

\begin{table}
 \caption{Average error (mas) for  COG, constrained matched filter and ANN results in the presence of turbulence for 316 and 1000 photons.}
 \label{symbols}
 \begin{tabular}{@{}lcccccc}
  \hline
  Case  & 1 & 2 & 3 & 4 \\
  \hline
  COG & 224.24 & 224.24 & 222.94 & 222.43 \\
  CMF & 143.94 & 135.00 & 224.06 & 220.29 \\
  ANN & 55.85 & 50.12 & 117.86 & 114.27 \\
  \hline
 \end{tabular}
 \medskip
 \\Case 1 is polar coordinate detector with 316 photons.\\
 Case 2 is polar coordinate detector with 1000 photons.\\
 Case 3 is cartesian CCD with diagonal spot elongation with 316 photons.\\
 Case 4 is cartesian CCD with diagonal spot elongation with 1000 photons.
 \label{table:ressummary}
\end{table}

\begin{table}
 \caption{Average error relative to COG error for constrained matched filter and ANN results in the presence of turbulence for 316 and 1000 photons}
 \label{symbols}
 \begin{tabular}{@{}lcccccc}
  \hline
  Case  & 1 & 2 & 3 & 4 \\
  \hline
  COG & 100 \% & 100 \% & 100 \% & 100 \% \\
  CMF & 64.19 \% & 60.20 \% & 100.50 \% & 99.03 \% \\
  ANN & 24.91 \% & 22.35 \% & 52.87 \% & 51.37 \% \\
  \hline
 \end{tabular}
 \medskip
 \\Case 1 is polar coordinate detector with 316 photons.\\
 Case 2 is polar coordinate detector with 1000 photons.\\
 Case 3 is cartesian CCD with diagonal spot elongation with 316 photons.\\
 Case 4 is cartesian CCD with diagonal spot elongation with 1000 photons.
 \label{table:ressummarypercent}
\end{table}

\subsection{Test run results}

\begin{figure}
  \includegraphics[width=84mm]{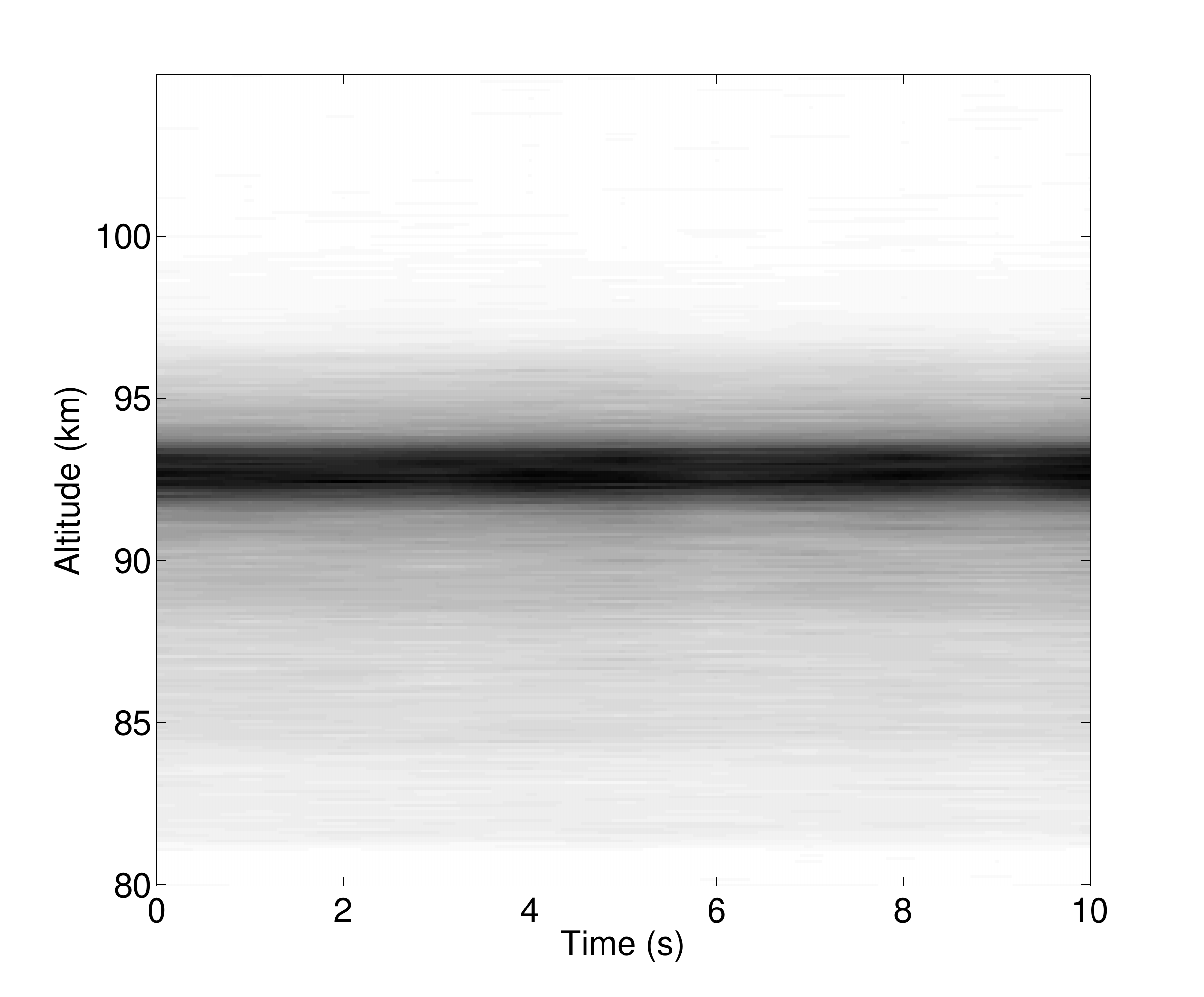}
  \caption{Continuously evolving sodium layer profile used in the validation tests.}
  \label{figure:contprofile}
\end{figure}

In this validation a portion of the measured profiles over a few seconds is chosen and interpolated to generate a continuous profile for a system working at 700$\,$Hz. This generates a real profile but with interpolation to achieve the desired rate. A phase screen following Kolmogorov statistics simulates the distorted wavefront to create realistic and time-dependent spot movements. Fig. \ref{figure:contprofile} shows the continuous profile used for this simulation.

The variables for the validation simulation were chosen to reflect the situation that would be experienced at the TMT telescope: a telescope diameter  of 30 metres using a Shack-Hartmann wavefront sensor  with 0.5 metres subapertures. Each subaperture images the artificial star on a CCD with 16x16 pixels. Kolmogorov phase-screens with $r_{0}$ =  0.15$\,$m were used. The wind speed determining the phase-screen motion was set at 10$\,\rmn{m s^{-1}}$. The photon noise level for this simulation was 1000 photons.

\begin{figure}
  \includegraphics[width=84mm]{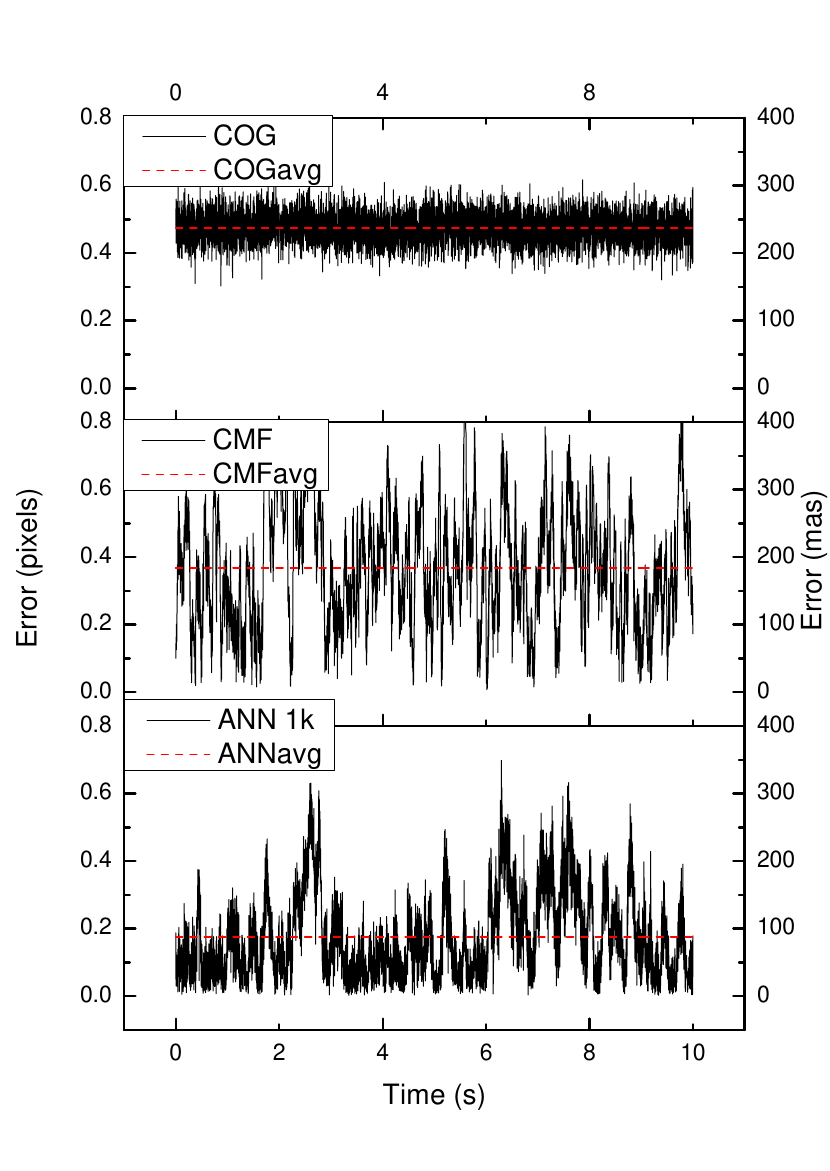}
  \caption{Error with evolving sodium layer and turbulence in a polar coordinate detector. The average value for each technique is shown as a dashed line. The averages are 0.474 pixels for COG, 0.368 pixels for CMF and 0.176 pixels for ANN.}
  \label{figure:validateturb}
\end{figure}

\begin{figure}
  \includegraphics[width=84mm]{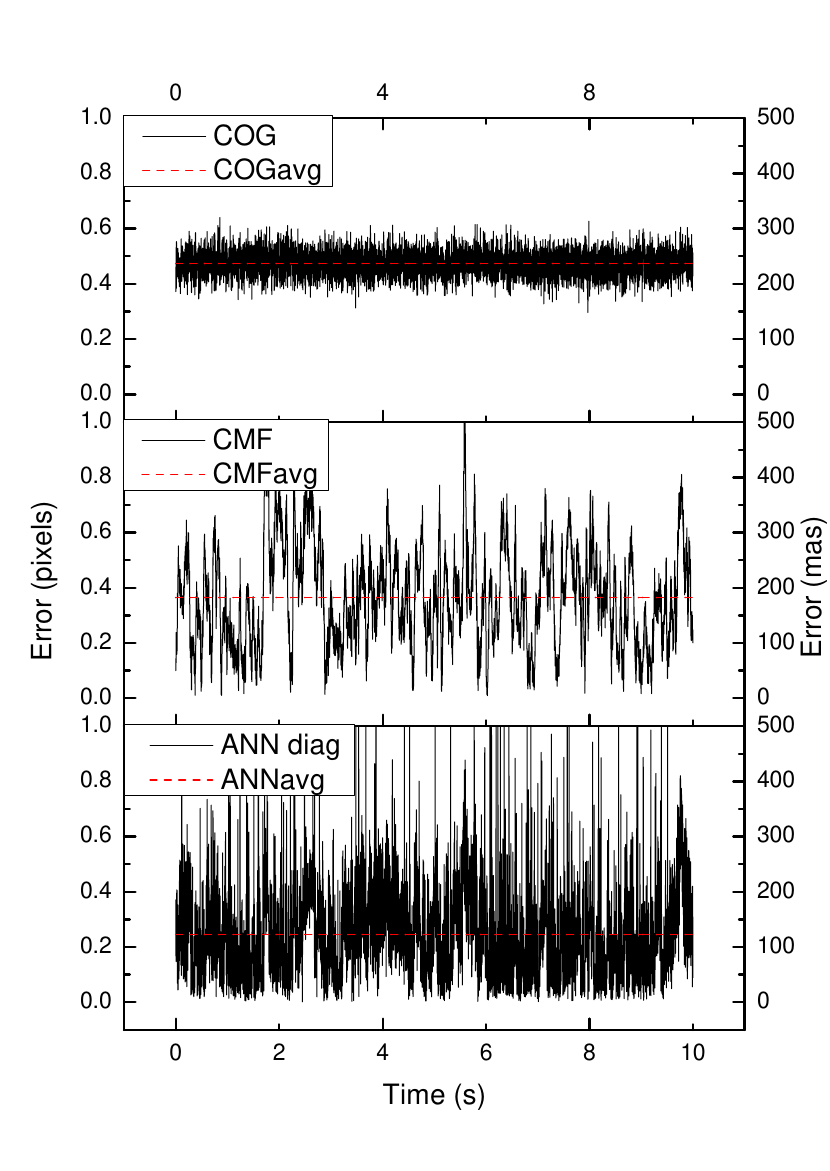}
  \caption{Error with evolving sodium layer and turbulence, cartesian CCD and diagonal elongation. Also shown the average value for each technique. The averages are 0.473 pixels for COG, 0.365 pixels for CMF and 0.245 pixels for ANN.}
  \label{figure:validatediag}
\end{figure}

Fig. \ref{figure:validateturb} shows a test executed with the evolving profile, a polar coordinate detector in the presence of turbulence. Fig. \ref{figure:validatediag} shows a test executed with the evolving profile and the presence of turbulence, and with a cartesian CCD using a spot with diagonal elongation as shown in Fig. \ref{figure:diagSubap}. The average results are also given in Table \ref{table:testruntable}. 

\begin{table}
 \caption{Average error (mas) for COG, constrained matched filter and ANN results for polar coordinate detector and cartesian CCD.}
 \label{symbols}
 \begin{tabular}{@{}lcccccc}
  \hline
    & COG & CMF & ANN \\
  \hline
  Polar & 237.31 & 184.23 & 88.06 \\
  Cartesian & 236.82 & 182.48 & 122.78 \\
  \hline
 \end{tabular}
 \medskip
 \label{table:testruntable}
\end{table}

The results show that the ANN performs better than other techniques in a more realistic simulation given an evolving sodium profile density and turbulence. In some moments both the ANN and the Matched Filter have a big error spikes, but both also have average errors below that of the centre of gravity technique, as expected. In a telescope using an adaptive optics system the science image will generally be a long exposure, so the average error is the determinant factor in image quality. 
But it can nonetheless create artefacts in the images so its use in high contrast imaging needs to be evaluated.

\section{Conclusions}

We have shown that in the presence of turbulence the ANN method for centroiding is superior and that the ANN is a viable and noise resistant technique for use in Shack-Hartmann wavefront sensors. Another advantage over other methods is that the ANN requires no calibration or reference at execution time.

The ANN method also has an advantage in open loop because it can handle big spot displacements if trained for it.  Furthermore, it does not require dithering, which is more difficult to implement in open loop and requires constant updating. In this work we have determined the following for ANN training: using noise in the training set generates better results for high noise, however the ANN will not give much better results if the noise level becomes less than the one used in training. In training we need to use images that are representative of reality, without being necessarily real data. Training with well simulated data gives good results. 

Another advantage of the ANN method is that it still works for conventional cartesian CCD.
For the spot elongation we had to train the ANN with the elongation direction that the ANN would see in reality, which means the use of a large number of different ANNs for a cartesian CCD, as each subaperture would see a different orientation. Future work will be directed at using larger data sets and improved ANN topology to obtain an ANN that can cope with spot direction variabilities. Validation tests with real data on an optical bench are also being planned.

\section*{Acknowledgements}

This project received financial support from FAPESC under project number TR2012-000034.
ATM appreciates support from CAPES and CNPq.
AK acknowledges support from CNPq.
DG appreciates support from FONDECYT through grant 11110149.
AG thanks support from FONDECYT, grant 1120626.
We appreciate Paul Hickson's support in providing LIDAR data.
This work has made use of the computing facilities of the Laboratory of Astroinformatics (IAG/USP).
ATM and AK appreciate support from INCT-A.

\label{lastpage}

\end{document}